\documentclass{article}
\pdfoutput=1

\usepackage{ijcai24}

\usepackage{algorithm}
\usepackage[noend]{algpseudocode}

\makeatletter
\@ifundefined{iffull}{\let\iffull\iftrue}{}
\@ifundefined{ifwarning}{\let\ifwarning\iffalse}{}
\makeatother

\newcommand{\avecsans}[2]{\iffull{#1}\else{#2}\fi}
\newcommand{\cfcompanion}{\avecsans{}{ in the extended paper}}
\usepackage{float}

\usepackage{environ}
\usepackage[title]{appendix}
\let\oriwrite=\write
\NewEnviron{myAppendix}{%
	\begin{appendices}%
		\iffull
		\BODY
		\else
		\ifwarning
		\else
		\setbox0=\vbox{\protected\def\write{\immediate\oriwrite}{\BODY}}
		\fi
		\fi
	\end{appendices}
}

\usepackage{times}
\usepackage{soul}
\usepackage{url}
\usepackage[hidelinks]{hyperref}
\usepackage[utf8]{inputenc}
\usepackage[small]{caption}
\usepackage{graphicx}
\usepackage{amsmath}
\usepackage{amssymb}
\usepackage{amsthm}
\usepackage{bbold}
\usepackage{booktabs}
\usepackage[switch]{lineno}

\usepackage{enumitem}

\usepackage{cleveref}

\usepackage{microtype}      
\usepackage{xcolor}         
\usepackage{icomma}
\usepackage{siunitx}

\usepackage[caption=false]{subfig}

\usepackage{tikz}
\usepackage{pgfplots}
\pgfplotsset{compat=1.18}
\tikzset{obj/.style={draw, circle, minimum size=.5cm}}
\tikzset{select/.style={red}}
\tikzset{placed/.style={font=\bfseries, very thick}}
\tikzset{median/.style={font=\bfseries, very thick, double}}
\tikzset{>=latex}

\newcommand{\execAsort}[3]{
	\begin{tikzpicture}
		\foreach \value/\style [count=\x] in {#1}
		{
			\node[obj, \style] (\value) at (\x, 0) {\strut{}\value};
		}
		\foreach \other in {#3}
		{
			\draw[-] (#2.north) to[bend left=15] (\other.north);
		}
	\end{tikzpicture}%
}

\newcommand{\execMerge}[3]{
	\begin{tikzpicture}
		\foreach \value [count=\x] in {#1}
		{
			\node[obj, anchor=center] (\value) at (\x, 0) {\strut{}\value};
		}
		\foreach \node/\other  in {#2}
		{
			\draw[-] (\node.north) to[bend left=30] (\other.north);
		}
		\foreach \left/\middle/\right in {#3}
		{
			\draw[black] (\left + .55, 0) -- ++(0, .4) -- ++(.1, 0);
			\draw[black] (\left + .55, 0) -- ++(0, -.4) -- ++(.1, 0);
			\draw[black] (\middle + .5, 0) -- +(0, .4) -- +(0, -.4);
			\draw[black] (\right + .45, 0) -- ++(0, .4) -- ++(-.1, 0);
			\draw[black] (\right + .45, 0) -- ++(0, -.4) -- ++(-.1, 0);
		}
	\end{tikzpicture}%
}

\usepackage[colorinlistoftodos,textsize=tiny]{todonotes}

\makeatletter
\renewcommand{\todo}[2][]{\tikzexternaldisable\@todo[#1]{#2}\tikzexternalenable}
\makeatother

\DeclareMathOperator*{\argmin}{arg\,min}

\title{\avecsans{Anytime Sorting Algorithms\\(Extended Version)}{Anytime Sorting Algorithms}}

\author{
Emma Caizergues$^{1, 2}$
\and
François Durand$^1$\and
Fabien Mathieu$^3$\\
\affiliations
$^1$Nokia Bell Labs, Massy, France\\
$^2$CNRS - Lamsade, Université Paris Dauphine - PSL, Paris, France\\
$^3$Swapcard Lab, Paris, France\\
}

\begin{document}

\maketitle

\begin{abstract}
This paper addresses the anytime sorting problem, aiming to develop algorithms providing tentative estimates of the sorted list at each execution step. Comparisons are treated as steps, and the Spearman's footrule metric evaluates estimation accuracy. We propose a general approach for making any sorting algorithm anytime and introduce two new algorithms: multizip sort and Corsort. Simulations showcase the superior performance of both algorithms compared to existing methods. Multizip sort keeps a low global complexity, while Corsort produces intermediate estimates surpassing previous algorithms.
\end{abstract}

\vspace{-.2cm}
\section{Introduction}\label{sec_contexte}

\paragraph{Motivation}

The objective of anytime sorting algorithms is to yield accurate estimates of a list's sorted order with a limited number of comparisons. This problem emerges particularly when the cost of comparing two items outweighs other computational operations.

For example, in scenarios involving human interaction, assigning a utility to each element of the list is not a reliable technique and it is better to proceed by comparing pairs~\cite{giesen2009approximate}. To illustrate this, consider a supervised learning based on sorted lists (e.g. learning to assert the quality of translations). Providing sorted lists instead of scores is relevant because humans are not good at grading many items, whereas pairwise comparisons are relatively accurate.
However, human comparisons are considerably more time-consuming than the other operations, which are performed by computers. To optimize efficiency, it is then crucial to distinguish comparison complexity from global complexity. In this context, anytime sorting has two main applications: instead of performing exact sorting on a small part of the raw input, one may wish to sort approximately more inputs. Uncertainty on the input and processing rates make it desirable to have an anytime approach. Also, users may be worn out after a certain amount of questions and want to stop the sorting process prematurely, so it is essential for the algorithm to generate accurate intermediate estimates that closely approximate the final sorted list.

Similar problems arise in situations where the considered items involve massive datasets requiring data transfers for each comparison~\cite{mesrikhani2018progressive}. The expense associated with each comparison necessitates the exploration of efficient algorithms that minimize the number of comparisons. Furthermore, practical constraints may require the sorting process to be interrupted before the algorithm's termination, so we need a reliable and accurate estimate of the sorted order based on the progress made thus far.

\paragraph{Related work}

Limiting the number of comparisons when sorting is a well-known problem~\cite{cormen}.
Among many famous algorithms such as quicksort, mergesort, heapsort, binary insertion sort and Shellsort, the Ford-Johnson algorithm~\cite{Ford1959ATP} stands out as very close to the theoretical limit in number of comparisons, and even optimal for certain values~\cite{peczarski2004new}. These algorithms do not consider a possible premature interruption.

A related problem is that of sorting under partial information. This problem entails determining a total order that is partially known, in the sense that a compatible partial order is available~\cite{supi}. Many variants of the sorting problem can be found within the realm of partial information, including scenarios such as ranking the $k$ best values~\cite{dushkin2018top}. While the partial information framework does not consider intermediate estimates, similarities with our contributions exist as we propose an algorithm that exploits the partial information extracted so far to determine the next comparison. Furthermore, partial information relates to classical sorts when considering empty information. For example, initializing Algorithm 1 in~\cite{supi} with no prior  amounts to the classical binary insertion sort, and their Algorithms 2 and 3 are both equivalent to mergesort (in its bottom-up version, which we discuss further in this paper).

Sorting when the probabilistic distribution of the inputs is known is another relevant problem to consider. In~\cite{moran2016note}, a variant of insertion sort is proposed as a solution, which has also been employed in~\cite{peters2021preference} for sorting with human comparisons. However, our approach differs in two main aspects. Firstly, our algorithms focus on providing good intermediate estimates throughout the sorting process. Secondly, we assume equal probabilities for all input permutations. In that case, their algorithm is equivalent to binary insertion sort. 

The potential interruption in the sorting process places us within the domain of anytime algorithms, which maintain result estimates at all times. Surprisingly, sorting has received limited attention in this literature, with studies focusing on specific algorithms like selection sort, Shellsort, and quicksort~\cite{grass1995programming,horvitz1988reasoning}. However, the metrics used to measure the estimation error are not always true distances.
In our benchmark, we exclude selection sort due to its $O(n^2)$ comparison complexity, while including Shellsort and quicksort for evaluation.

Among related problems, \emph{progressive algorithms} can also be interrupted at any time, but the focus is on provable bounds for worst-case performance rather than on empirical average efficiency~\cite{alewijnse2014progressive}. Contract-based algorithms also make a trade-off between time and accuracy, but assume that the available time is known in advance~\cite{zilberstein1995optimizing,zilberstein1996using}. 

One of the most closely related studies is the work by \cite{giesen2009approximate}, which delves into the link between the number of comparisons performed and the estimation error. Using Spearman's footrule metric to evaluate the error, they show that, for a list of size $n$, the error after $k\leq n(\ln(n)-6)$ comparisons is at least $n^2e^{-k/n-6}$ in the worst case. Conversely, they introduce ASort, an anytime algorithm with guarantee that the error is less than $n^2e^{-k/6n+1}$. While these results show that ASort is asymptotically optimal in some sense, the ratio between the two bounds above ($e^{5k/6n+7}$) is at least $e^7>1000$ and potentially unbounded. This means that the practical performance of ASort remains to be investigated. One of the goals of our work is to provide a practical benchmark for ASort while also introducing new algorithms with superior empirical performance.

\paragraph{Contributions}
In this paper, we make the following contributions to the field of anytime sorting:
\begin{enumerate}[align=left, leftmargin=\parindent, labelindent=\parindent,
 listparindent=\parindent, labelwidth=0pt, itemindent=!]
	\itemsep0em 
    \item We generalize the work on ASort by \cite{giesen2009approximate} under the anytime framework and propose the anytime sorting problem, where the \emph{performance profile} of an algorithm is measured by the \emph{Spearman's footrule metric} between its tentative estimates and the perfectly sorted list, as a function of the number of comparisons already performed.
    \item We reexamine classical sorting algorithms from the perspective of anytime sorting.
    \item We propose simple heuristics called \emph{estimators}, which leverage the available partial information at each step to approximate the final sorted result.
    \item We present an enhanced version of the traditional mergesort algorithm called \emph{multizip sort}. This algorithm is specifically designed to improve the anytime sorting capabilities.
    \item We introduce the \emph{Corsort} family of anytime sorting algorithms, which rely on estimators to determine the next comparison to perform.
    \item Using extensive simulations, we analyze the performance of Corsort, multizip, and ASort.
    \begin{itemize}[align=left, leftmargin=\parindent, labelindent=\parindent,
 listparindent=\parindent, labelwidth=0pt, itemindent=!]
    \itemsep0em 
        \item Corsort has a quasi-optimal comparison complexity and provides superior intermediate estimates compared to existing state-of-the-art sorting algorithms. Its main drawback is its global complexity.
        \item Multizip sort has a better comparison complexity. It maintains  in most cases the second-best intermediate results. Its main drawback is its interruption cost.
        \item ASort has less optimal comparison complexity and performance profile. Its main benefits are low global complexity and absence of interruption cost.
    \end{itemize}
\end{enumerate}
A comprehensive summary of the computational complexities is provided in Table \ref{tab:complexity}.

\begin{table*}[!ht]
	\centering

 \begin{tabular}{lllll}
 \hline
	& Comparison & Global & Native & Interest(s) \\
	& complexity & complexity & estimator &  \\
	\hline
Ford-Johnson & $\sim n\log_2(n)$ & $O(n^2)$ & No & Comparison complexity  \\
Binary insertion & $\sim n\log_2(n)$ & $O(n^2)$ & Yes & Comparison complexity \\
Quicksort & $O(n\ln(n))$ & $O(n\ln(n))$ & Yes & Fast termination in practice \\
ASort & $O(n\ln(n))$ & $O(n\ln(n))$ & Yes & Good native estimator with theoretical guarantees\\
Top-down merge & $\sim n\log_2(n)$ & $O(n\ln(n))$ & Yes & Complexities  \\
Bottom-up merge & $\sim n\log_2(n)$ & $O(n\ln(n))$ & Yes  & Good performance profile, complexities \\
Multizip & $\sim n\log_2(n)$ & $O(n\ln(n))$ & Yes  & Very good performance profile, complexities \\
Corsort & $O(n^2)^*$ & $O(n^4)^*$ & No & Best performance profile, comparison complexity$^*$\\
	\hline
\end{tabular}
\newline
\vspace{-.1cm}
\caption{Cost of the algorithms considered in this paper (we exclude heapsort and Shellsort due to their poor termination times, see \Cref{fig:total}). All complexities are well-known results except for the last two rows. All algorithms can use the estimator $\rho$ when interrupted. Using $\rho$ requires $O(n\ln(n))$ for Corsort, $O(n^3)$ operations for the other algorithms. Native estimators, when they exist, can be used directly.\\\hspace{\textwidth}
$^*$ Simulations suggest $O(n\ln(n))$ comparisons, yielding a global complexity of $O(n^3\ln(n))$ operations.
\label{tab:complexity}
}
\vspace{-.4cm}
\end{table*}

\paragraph{Limitations}
The anytime sorting algorithms we propose are designed to address specific scenarios, and their suitability depends on the following assumptions:
\begin{enumerate}[align=left, leftmargin=\parindent, labelindent=\parindent,
 listparindent=\parindent, labelwidth=0pt, itemindent=!]
	\itemsep0em 
    \item It is impractical or unreliable to assign numerical utilities to the items to sort.
    \item The sorting process may be interrupted due to tight time constraints or resource limitations.
    \item An approximate ranking is acceptable, although the approximation error does matter.
    \item Good empirical performance is required, possibly at the expense of formal guarantees.
    \item The cost of comparisons is orders of magnitude higher than other computing operations.
    \item Worse case scenarios and non-uniform dataset/inputs are not the point of the study.
\end{enumerate}

If these assumptions do not hold, alternative solutions may be more appropriate. For example, algorithms like Ford-Johnson are well-suited for situations with costly comparisons and a low likelihood of interruption; contract-based algorithms offer better suitability when interruption times are known and formal guarantees are required. ASort offers similar guarantees, even though it is outperformed experimentally.

Furthermore, our proposal does not rely on any specific prior information about the input list. This makes our proposal very generic but also means that if such prior exists, we do not leverage it to reduce the average number of comparisons required, as it is done in \cite{peters2021preference}. 

Lastly, we have not provided a formal proof that our Corsort algorithm has an average comparison complexity in $O(n\ln(n))$. This conjecture is only supported by simulations demonstrating its superior performance compared to heapsort, Shellsort, and quicksort/ASort.

\paragraph{Roadmap}
The rest of the article is organized as follows. \Cref{sec_estimateurs} formally defines anytime sorting, presents a taxonomy of anytime sorting algorithms, and introduces our novel approaches: estimators, multizip sort, and Corsort. \Cref{sec_eval} evaluates the proposed solutions through simulations. \Cref{sec_conclu} concludes.
\avecsans{%
\paragraph{Published version} The present paper is an extended version of \cite{cdm24ijcai}.
}{%
\paragraph{Extended version} A version of the paper completed by an Appendix is available online \cite{cdm24extend}.
}

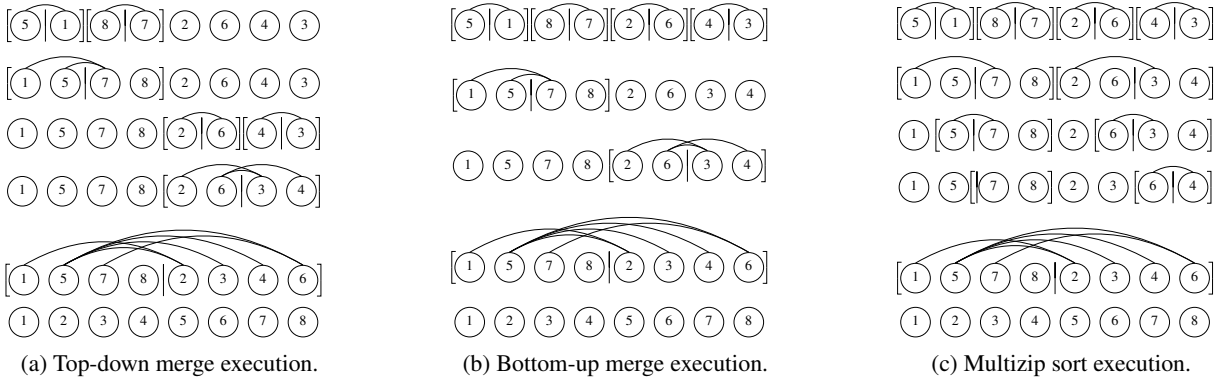
\begin{figure*}[!ht]
	\hfill
	\subfloat[Top-down merge execution.\label{fig:mergesort_exec}]{%
\begin{tikzpicture}[scale=.53, transform shape]
	\def\interspace{.02cm}
	\node (x0) {\execMerge{5, 1, 8, 7, 2, 6, 4, 3}{5/1, 8/7}{0/1/2, 2/3/4}};
	\node[below = \interspace of x0] (x1) {\execMerge{1, 5, 7, 8, 2, 6, 4, 3}{1/7, 5/7}{0/2/4}};
	\node[below = \interspace of x1] (x2) {\execMerge{1, 5, 7, 8, 2, 6, 4, 3}{2/6, 4/3}{4/5/6, 6/7/8}};
	\node[below = \interspace of x2] (x3) {\execMerge{1, 5, 7, 8, 2, 6, 3, 4}{2/3, 6/3, 6/4}{4/6/8}};
	\node[below = \interspace of x3] (x4) {\execMerge{1, 5, 7, 8, 2, 3, 4, 6}{1/2, 5/2, 5/3, 5/4, 5/6, 7/6}{0/4/8}};
	\node[below = \interspace of x4] (x5) {\execMerge{1, 2, 3, 4, 5, 6, 7, 8}{}{}};
\end{tikzpicture}
	}	
	\hfill
	\subfloat[Bottom-up merge execution.\label{fig:bottom_up_exec}]{%
\begin{tikzpicture}[scale=.53, transform shape]
	\def\interspace{.34cm}
	\node (x0) {\execMerge{5, 1, 8, 7, 2, 6, 4, 3}{5/1, 8/7, 2/6, 4/3}{0/1/2, 2/3/4, 4/5/6, 6/7/8}};
	\node[below = \interspace of x0] (x1) {\execMerge{1, 5, 7, 8, 2, 6, 3, 4}{1/7, 5/7}{0/2/4}};
	\node[below = \interspace of x1] (x2) {\execMerge{1, 5, 7, 8, 2, 6, 3, 4}{2/3, 6/3, 6/4}{4/6/8}};
	\node[below = \interspace of x2] (x3) {\execMerge{1, 5, 7, 8, 2, 3, 4, 6}{1/2, 5/2, 5/3, 5/4, 5/6, 7/6}{0/4/8}};
	\node[below = \interspace of x3] (x4) {\execMerge{1, 2, 3, 4, 5, 6, 7, 8}{}{}};
\end{tikzpicture}
	}	
	\hfill
	\subfloat[Multizip sort execution.\label{fig:multizip_exec}]{%
\begin{tikzpicture}[scale=.53, transform shape]
	\def\interspace{.07cm}
	\node (x0) {\execMerge{5, 1, 8, 7, 2, 6, 4, 3}{5/1, 8/7, 2/6, 4/3}{0/1/2, 2/3/4, 4/5/6, 6/7/8}};
	\node[below = \interspace of x0] (x1) {\execMerge{1, 5, 7, 8, 2, 6, 3, 4}{1/7, 2/3}{0/2/4, 4/6/8}};
	 \node[below = \interspace of x1] (x2) {\execMerge{1, 5, 7, 8, 2, 6, 3, 4}{5/7, 6/3}{1/2/4, 5/6/8}};
	\node[below = \interspace of x2] (x3) {\execMerge{1, 5, 7, 8, 2, 3, 6, 4}{6/4}{1.88/2.08/4, 6/7/8}};
	\node[below = \interspace of x3] (x4) {\execMerge{1, 5, 7, 8, 2, 3, 4, 6}{1/2, 5/2, 5/3, 5/4, 5/6, 7/6}{0/4/8}};
	\node[below = \interspace of x4] (x5) {\execMerge{1, 2, 3, 4, 5, 6, 7, 8}{}{}};
\end{tikzpicture}
	}	
	\hfill
	\caption{Sorting the list $X=(51872643)$ with top-down merge, bottom-up merge and multizip sort. Each edge represents a comparison. The bracket notation $[\;\mid\;]$ delimits two sublists already sorted that are being merged. For more details, cf. Appendix~\ref{sec:detailed_execution_multizip}\cfcompanion.\label{fig:mergeexecutions}}
	\vspace{-.3cm}
\end{figure*}

\section{Anytime sorting}\label{sec_estimateurs}

Formally, we want to sort a list $X = (X[1], \ldots, X[n])$, where $n > 0$, by performing comparisons of the type: is $X[i]$ less than $X[j]$? An \emph{anytime sorting algorithm} is an algorithm capable, after $k$ comparisons have been performed, of returning an estimate $X_k$ of the result. By convention, if $k$ is greater than the number of comparisons needed for the algorithm's termination, then $X_k$ is the sorted list. We measure the error of $X_k$ by computing the Spearman's footrule metric between $X_k$ and the sorted list~\cite{diaconis1977spearman}, i.e. the sum of the absolute differences between the ranks of elements in $X_k$ and in the sorted list : $S_k = \sum_{i}| r(X_k[i])-i|$, where $r(x)$ denotes the rank of $x$ in the sorted list.
The function $k \to S_k$, which represents the evolution of the error made as the algorithm runs, is called its \emph{performance profile}. Ideally, we are looking for an anytime sorting algorithm whose performance profile is consistently lower than that of the other algorithms tested.

\subsection{Classical sorting}\label{sec:tris_classiques}

Some classical algorithms can be seen as anytime because they maintain a list that converges to the sorted list and can be used as an estimate $X_k$. This is the case of quicksort, mergesort, binary insertion sort, and Shellsort, all of which we shall implement in a way that is favorable to the spirit of the original algorithm: for quicksort, for instance, the position of the pivot is updated in the list after each comparison.

Changing the order in which comparisons are made may improve the intermediate estimates $X_k$. For mergesort, the standard recursive implementation, also called \emph{top-down}, sorts the left part, then the right part of each considered sub-list (recursively), and reunites them by a merge procedure~\cite[Chapter~2]{cormen}. However, it is also possible to go through it \emph{bottom-up}, by sorting sub-lists of increasing size~\cite[Chapter 5.2.4]{knuth1997art}. Figure~\ref{fig:mergeexecutions} illustrates this difference on an example\footnote{For brevity, we omit the commas to note the lists taken as examples in the tables and figures. For example, the list $(a, b, c, d)$ is noted as $(abcd)$.} of size $8$: the top-down execution sorts completely the left side of the list before completely sorting the right side of the list, whereas the bottom-up execution starts by sorting all lists of size 2, both on left and right sides. In all classical mergesort implementations, the merge procedure is a single block.
We propose a variant of the bottom-up implementation, which we call \emph{multizip sort}, where all merge procedures of a given depth of the recursion tree are interleaved. Figure~\ref{fig:multizip_exec} illustrates the idea: the first part of the execution is unchanged and sorts all lists of size $2$. Then, when \emph{bottom-up} sorts the two lists of size $4$ sequentially, \emph{multizip} alternates comparisons in the two lists until they are both sorted (cf. Appendix~\ref{sec:pseudocode_multizip}\cfcompanion{} for details).
The rationale for proposing \emph{multizip sort} will be discussed in Section~\ref{subsec:estimators}. Note that all the considered variants of mergesort make the exact same comparisons, but not in the same order. Intuitively, bottom-up and multizip ensure that the information available on the elements of the list (through comparisons) is more balanced. This should be more favorable for the intermediate estimates~$X_k$, as \Cref{sec:profiles} will verify experimentally.

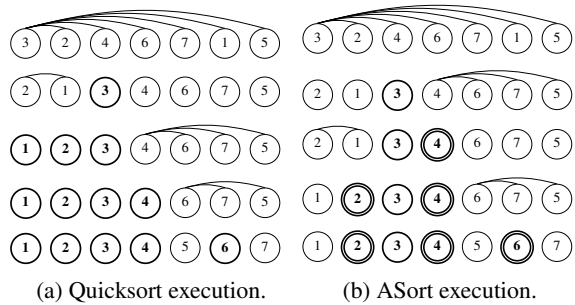
\begin{figure}[!ht]
	\vspace{-.45cm}
\centering
\subfloat[Quicksort execution.\label{fig:quicksort_exec}]{%
		\begin{tikzpicture}[scale=.53, transform shape]
	\def\interspace{.1cm}
	\node (x0) {\execAsort{3/, 2/, 4/, 6/, 7/, 1/, 5/}{3}{2, 4, 6, 7, 1, 5}};
	\node[below = \interspace of x0] (x1) {\execAsort{2/, 1/, 3/placed, 4/, 6/, 7/, 5/}{2}{1}};
	\node[below = \interspace of x1] (x3) {\execAsort{1/placed, 2/placed, 3/placed, 4/, 6/, 7/, 5/}{4}{6, 7, 5}};
	\node[below = \interspace of x3] (x4) {\execAsort{1/placed, 2/placed, 3/placed, 4/placed, 6/, 7/, 5/}{6}{7, 5}};
	\node[below = \interspace of x4] (x5) {\execAsort{1/placed, 2/placed, 3/placed, 4/placed, 5/, 6/placed, 7/}{1}{}};
\end{tikzpicture}
	} 
 \subfloat[ASort execution. \label{fig:asort_exec}]{%
		\begin{tikzpicture}[scale=.53, transform shape]
	\def\interspace{.1cm}
	\node (x0) {\execAsort{3/, 2/, 4/, 6/, 7/, 1/, 5/}{3}{2, 4, 6, 7, 1, 5}};
	\node[below = \interspace of x0] (x1) {\execAsort{2/, 1/, 3/placed, 4/, 6/, 7/, 5/}{4}{6, 7, 5}};
	\node[below = \interspace of x1] (x2) {\execAsort{2/, 1/, 3/placed, 4/median, 6/, 7/, 5/}{2}{1}};
	\node[below = \interspace of x2] (x3) {\execAsort{1/, 2/median, 3/placed, 4/median, 6/, 7/, 5/}{6}{7, 5}};
	\node[below = \interspace of x3] (x4) {\execAsort{1/, 2/median, 3/placed, 4/median, 5/, 6/median, 7/}{1}{}};
\end{tikzpicture}
	}
	\caption{Sorting the list $X=(3246715)$ with quicksort or ASort.
 A node appears in bold if it has already been used as a pivot: it partitions the list into smaller elements on the left and larger ones on the right.
 Intermediate steps that do not perform any comparison are omitted. 
 For ASort (\ref{fig:asort_exec}), we use quickselect~\protect\cite{hoare1961algorithm} as median identification subroutine. Each step represents the application of a pivot, and each edge represents a comparison. A node is circled twice if it has been identified as a median. In the third step, since the median 4 has been found, we must compute the median of the left sub-list $(213)$; but note that it is useless to make any comparison with element 3, previously used as a pivot, because it is is already in its final position.
 \label{fig:quickexecutions}
 }
\end{figure}

For quicksort, balancing the comparisons is equivalent to the ASort algorithm \cite{giesen2009approximate}, using quickselect~\cite{hoare1961algorithm} as a subroutine for median identification. Specifically, ASort determines the median, separates smaller from larger elements, and proceeds recursively. When utilizing quickselect to identify the median, which uses successive pivots like quicksort, ASort performs precisely the same comparisons as quicksort, albeit not in the same order. This is illustrated in \Cref{fig:quickexecutions}.

Certain other classical algorithms allow to obtain an estimate $X_k$ by a natural transformation of their current state. For example, heapsort keeps in memory the heap associated to the partially sorted list. To obtain a fair estimate from the algorithm, we can first go through the heap (backwards), then through the elements already sorted (forwards).

Finally, some algorithms like Ford-Johnson, whose internal state does not have a structure naturally close to a list,  do not seem to have a natural estimator.
Since it is not always trivial to translate the execution of an algorithm into intermediate estimators, we show in the next section how to produce $X_k$ for any comparison-based sorting algorithm.

\subsection{Estimators}
\label{subsec:estimators}

To make an anytime version of any sort, we propose to use an estimator that ignores the execution details and solely relies on the historical record of the comparisons made.

Let $C_k=\{X[i_1]<X[j_1], \ldots, X[i_k]< X[j_k]\}$ be the result of $k$ comparisons. $C_k$ defines a partial order $\preceq_k$ on the elements of the list by transitive closure\footnote{We assume for clarity that all elements are distinct. In case of redundancy, uniqueness can be enforced by appending to each element $x$ its index $i$ in the initial list. For example, $(17, 42, 42)$ should be seen as $((17, 1), (42, 2), (42, 3))$. When sorting the list, it is possible to be in a situation where we already know that $(17, 1) \preceq_k (42, 2)$ but we do not yet know how to compare $(17, 1)$ and $(42, 3)$.}. We call \emph{estimator} a function that associates to any partial order one of its \emph{linear extensions}, i.e. a compatible total order.

An estimator is optimal if it always finds a total order that minimizes the expected Spearman's footrule metric with a uniform random linear extension of the input partial order. Assuming the linear extensions of $\preceq_k$ are known, this is equivalent to solving an assignment problem~\cite{assignment55hungarian} where the cost of assigning element $x$ to index $i$ is the total error generated by the decision over all extensions. As the metric is a $L_1$-norm, we can also associate to each element its median rank, which gives the optimal result as long as all medians are distinct integers. However, both approaches are computationally challenging, as enumerating the linear extensions is known to be \#P-complete~\cite{brightwell1991counting}.

In practice, we propose to use classical \emph{score-and-sort} heuristics~\cite{calauzenes2020ranking,robertson1997probability}: each element gets a score that reflects its estimated ranking in the list, and we return the list associated with sorting the scores. Recall that global complexity is distinct from comparison complexity: we assume that sorting $n$~numerical scores is much faster than comparing two elements of the list. In case of ties, we return the elements in their original order. By a slight abuse of language, we also call \emph{estimator} the score function itself.

To build our heuristics, we define descendants and ancestors as follows: if $x$ is an element of the list,  $d_k(x)=|\{y\in X: y \preceq_k x\}|$ and $a_k(x)=|\{y\in X: x \preceq_k y\}|$ are respectively the number of known descendants and ancestors of $x$ in $\preceq_k$ ($x$ is included in both sets by convention). A simple way to compute $d_k$ and $a_k$ for a given $k$ is to build the transitive closure of $C_k$, which can be done in $O(n^3)$ using the Flyod-Marshall algorithm~\cite[chapter 25.2]{cormen}. If one wishes $d_k$ and $a_k$ for all values of $k$, it is best  to maintain the sets of descendants and ascendants: after each new comparison, update the descendants of the ancestors of the greater element with the descendants of the smaller element, and conversely (cf Appendix~\ref{sec:pseudocode_estimators}\cfcompanion{} for details). The cost of this incremental approach is $O(n^2)$ for each new comparison performed.

Using $d_k$ and $a_k$, we consider the following score functions:
\begin{itemize}[align=left, leftmargin=\parindent, labelindent=\parindent,
 listparindent=\parindent, labelwidth=0pt, itemindent=!]
	\itemsep0em 
	\item $\Delta_k$, defined by $\Delta_k(x)=d_k(x)-a_k(x)$. $\Delta_k$ assigns to each $x$ a score that reflects the average between its lowest and highest possible positions. 
	\item $\rho_k$, defined by $\rho_k(x)=d_k(x)/(d_k(x)+a_k(x))$. $\rho_k$ positions $x$ as if its descendants and ancestors were on average regularly spaced within the whole list. 
\end{itemize}

If there is no ambiguity, we will omit the index $k$.
$\Delta$, $\rho$, but also the median rank $m$ are valid estimators: one can check that if an element is greater than another in a partial order, then its score is higher, which ensures that the returned estimate is a linear extension. In particular, when the result of all comparisons is known, sorting the elements according to the score function returns the sorted list.

\begin{figure}[!ht]
	\centering
	\begin{tikzpicture}[scale=.45, transform shape]
\footnotesize
\foreach \label/\desc/\anc/\x/\m/\mn in {
  a/1/2/-1.5/5/0.2777777777777778,
  b/2/1/-1.5/13/0.7222222222222222,
  c/1/15/0/1/0.05555555555555555,
  d/2/14/0/2/0.1111111111111111,
  e/3/13/0/3/0.16666666666666666,
  f/4/12/0/4/0.2222222222222222,
  g/5/11/0/6/0.3333333333333333,
  h/6/10/0/7/0.3888888888888889,
  i/7/9/0/8/0.4444444444444444,
  j/8/8/0/9/0.5,
  k/9/7/0/10/0.5555555555555556,
  l/10/6/0/11/0.6111111111111112,
  m/11/5/0/12/0.6666666666666666,
  n/12/4/0/14/0.7777777777777778,
  o/13/1/-1/16/0.8888888888888888,
  p/13/2/1/15/0.8333333333333334,
  q/14/1/1/17/0.9444444444444444
}
{
  \pgfmathtruncatemacro{\to}{\desc + \anc}
  \pgfmathtruncatemacro{\delta}{\desc - \anc}
  \node[obj,inner sep=1pt,minimum size=1pt] (d\label) at (1.15*\x, 7.01+.624*13/15*\delta) {\Large\strut$\label$};
  \node[right = -.5mm of d\label] {\Large$\delta$};
  \node[obj,inner sep=1pt,minimum size=1pt] (r\label) at (7+1.4*\x, 16.5*\desc/\to - 1.5) {\Large\strut$\label$};
  \node[right = -.5mm of r\label] {\Large$\frac{\desc}{\to} $};
  \node[obj,inner sep=1pt,minimum size=1pt] (h\label) at (14+1.6*\x, 16.5*\mn - 1.5) {\Large\strut$\label$};
  \node[right = -.5mm of h\label] {\Large $\m$};
}
\node at (0, -1.4) {\LARGE(a) $\Delta$ scores.};
\node at (7, -1.4) {\LARGE(b) $\rho$ scores.};
\node at (14, -1.4) {\LARGE(c) $m$ scores.};
\foreach \i/\j in {a/b, c/d, d/e, e/f, f/g, g/h, h/i, i/j, j/k, k/l, l/m, m/n, n/o, n/p, p/q}{%
  \draw[<-] (d\i) -- (d\j);
  \draw[<-] (r\i) -- (r\j);
  \draw[<-] (h\i) -- (h\j);
}
\end{tikzpicture}
    \vspace{-.6cm}
\begin{tabular}{llS[table-format=2.2]}
		\hline
		Estimator & Returned estimate & $\overline{S}$ \\
		\hline
		$\Delta$ & $(cdefghiajbklmnpoq)$ & 16.2\\
		$\rho$ & $(cdefgahijklbmnpoq)$ & 14.0 \\
		$m$ & $(cdefaghijklmbnpoq)$ & 13.9\\
		\hline
\end{tabular}
	\caption{Example of score-based selection of a linear extension. The input partial order, represented by the edges of the graph, is the transitive closure of: $a\prec b$, $c\prec d \prec \ldots \prec n \prec o$, $n\prec p \prec q\text{.}$ Heights are proportional to scores. Returned estimates and expectation $\overline{S}$ of the error $S$ are also provided for completeness.
		\label{fig:differences_estimators}}
		\vspace{-.4cm}
\end{figure}

Figure \ref{fig:differences_estimators} exhibits a partial order where the three estimators give distinct results. Here, $\rho$ is better than $\Delta$ because it positions better the small chain component (nodes $a$ and $b$) with respect to the large Y-shaped one (nodes $c$ to $q$). However, it is still surpassed by the median rank~$m$, which is optimal here because all medians are distinct (cf. above).

Figure \ref{fig:differences_estimators} illustrates that $\Delta$ and $\rho$ distort their estimate when the partial order is a combination of chains and $Y$-shaped components. Yet, during a standard mergesort execution, a typical partial order is a combination of chains (the sorted sublists) and one $Y$-shaped component (the ongoing merge). We introduced the multizip sorting earlier to avoid this scenario, so that a typical partial order will be multiple $Y$-shaped components of similar size (the ongoing merge procedures). In essence, multizip is a variant of mergesort tailored for $\Delta$ and $\rho$.

\paragraph{Multizip complexity} As multizip schedules the same comparisons as mergesort, its comparison complexity is equivalent to $n\log_2(n)$ even in worst case, which is asymptotically optimal~\cite{cormen}. Each step requires $O(1)$ computing operations so the global complexity is $O(n\ln(n))$. If an interruption occurs, we need to build the transitive closure of the comparisons made, which is done in $O(n^3)$.

In our preliminary work \cite{corsort}, we found out that $\rho$ generally outperforms $\Delta$ as an estimator, and therefore we adopt it as the default choice. However, $\Delta$ can serve another purpose, as elaborated in \Cref{sec_Corsort}.

\subsection{Comparison-Oriented sort}\label{sec_Corsort}

We call \emph{Corsort} (\emph{Comparison-ORiented Sort}) a sorting algorithm that, at each step of its execution, selects the next comparison based on the current partial order $\preceq_k$. We assume that only pairs that are not comparable yet (according to $\preceq_k$) are chosen until convergence is achieved.

The heart of a Corsort, i.e., the choice of the next comparison, aims at two objectives. Firstly, it must ensure fast termination, i.e., minimize the total number of comparisons. This is a problem of \emph{sorting under partial information}, which amounts to choosing a comparison whose result is as uncertain as possible~\cite{supi}. To this end, we propose to rely on closeness for a score function. The rationale is that close scores should indicate a high uncertainty regarding the outcome of the comparison. Secondly, to improve the quality of intermediate estimates $X_k$, we need to acquire information on elements for which we have little, as these create uncertainty that will be reflected in the error $S_k$. We introduce $I_k(x) = d_k(x) + a_k(x)$ to represent the amount of ``information'' acquired on an element $x$, and we wish to compare elements for which $I_k$ is low. 
In our preliminary work \cite{corsort}, we compared several variants of Corsort and we selected the following one, called Corsort-$\Delta$:
\begin{itemize}[align=left, leftmargin=\parindent, labelindent=\parindent,
 listparindent=\parindent, labelwidth=0pt, itemindent=!]
	\itemsep0em 
	\item The next comparison is made by seeking a pair of incomparable items whose $\Delta$ scores are as close as possible.
	\item We use $I_k$ for tie-breaking. Specifically, among the pairs that minimize $\Delta$-closeness, we pick up one that minimizes $\max(I_k(x), I_k(y))$.
\end{itemize}

\begin{figure}[!t]
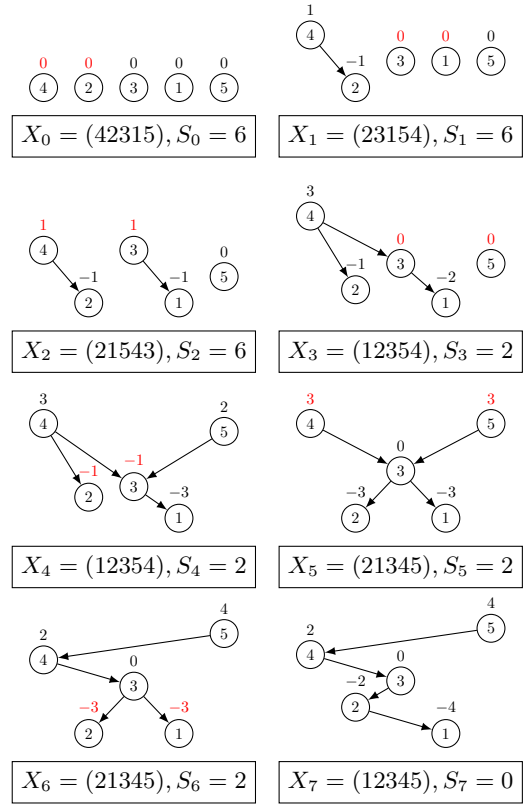

	\centering
	\include{corsort_execution}
	\vspace{-.5cm}
	\caption{Execution of Corsort-$\Delta$ on the list $X=(42315)$. Each step $k$ depicts the partial order after $k$ comparisons. The next comparison to perform is visually highlighted by the two vertices whose values of $\Delta_k$ (displayed above each element) are in red. Each subfigure represents $\rho_k$ as each element's height and gives the corresponding estimate $X_k$ based on $\rho_k$ and the associated error $S_k$.}
	\label{fig:Corsort_exec}
	\vspace{-.4cm}
\end{figure}

The intermediate estimates $X_k$ can be given by any estimator. In practice, we use  $\rho$. For the pseudocode, cf. Appendix \ref{sec:pseudocode_corsort}\cfcompanion. As an example, Figure \ref{fig:Corsort_exec} shows the execution of the selected Corsort on the list $X=(42315)$. 
As we argued earlier that $\rho$ is better than $\Delta$, one may wonder why $\Delta$ is preferred for the choice of the comparison. An intuition that explains the advantage of $\Delta$ is the need of balance between termination and intermediate estimates: $\rho$ can take $O(n^2)$ distinct values, against $O(n)$ for $\Delta$. As a result, $\Delta$ generates more ties than $\rho$ (a typical case is $X_4$ on Figure~\ref{fig:Corsort_exec}: $\rho_4(2)<\rho_4(3)$ but $\Delta_4(2)=\Delta_4(3)$), so the tie-breaking rule, which targets accuracy of the estimate, is used more often. Experimentally, this gives a better performance profile than a $\rho$-based Corsort, at the price of a (slightly) longer termination.

\paragraph{Complexity} Corsort never compares a given pair twice, so it makes at most $n(n-1)/2$ comparisons. At each step, one needs to compute the next pair, compare, then update $a$ and~$d$, for a cost of $O(n^2)$ operations, which sums up to a global complexity in $O(n^4)$. The cost of interruption, a score-and-sort in $O(n\ln(n))$,  is comparatively negligible.

\section{Evaluation}\label{sec_eval}

To compare the performance of ASort, multizip sort and Corsort with the classical sorting algorithms discussed in \Cref{sec:tris_classiques}, we have developed an open-source Python package specifically designed for creating anytime sorting algorithms and evaluating their effectiveness~\cite{corsort} (cf. Appendix~\ref{sec:implementation}\cfcompanion).

\subsection{Uninterrupted behavior}\label{sec:termination_time}

\begin{figure}[!t]
	\centering
	\begin{tikzpicture}

\definecolor{heap}{RGB}{170,170,170}
\definecolor{shell}{RGB}{176,176,176}
\definecolor{asort}{RGB}{214,39,40}
\definecolor{quicksort}{RGB}{200,200,14}
\definecolor{fj}{RGB}{111,54,29}
\definecolor{binary}{RGB}{255,100,193}
\definecolor{merge}{RGB}{31, 119, 180}
\definecolor{multizip}{RGB}{15,15,15}
\definecolor{corsort}{RGB}{44,200,44}

\definecolor{lightgray204}{RGB}{204,204,204}

\begin{axis}[width=8.8cm, height=7cm,
legend cell align={left},
legend columns = 4,
transpose legend,
legend style={fill opacity=0.8, draw opacity=1, text opacity=1, draw=lightgray204, nodes={scale=0.7, transform shape}, anchor=north east, at={(.99, .9)}},
log basis x={10},
tick align=outside,
x grid style={darkgray176},
xlabel={Size \(\displaystyle n\) of the list},
x label style={at={(0.5,-0.1)}},
xmin=8, xmax=2048,
xmode=log,
xtick pos=left,
xtick style={color=black},
y grid style={darkgray176},
ylabel={Relative overhead (\%)},
y label style={at={(-0.08,0.5)}},
ymin=0, ymax=100,
ytick pos=both,
ytick style={color=black}
]
\path [draw=heap, fill=heap, opacity=0.2]
(axis cs:8,83.195865430739)
--(axis cs:8,50.4823180323927)
--(axis cs:16,69.5197694840764)
--(axis cs:32,81.8807556649569)
--(axis cs:64,88.5178026451842)
--(axis cs:128,92.2752589937676)
--(axis cs:256,94.5373002573378)
--(axis cs:512,95.9658043342508)
--(axis cs:1024,96.8182020681088)
--(axis cs:2048,97.3781015120632)
--(axis cs:2048,98.2156829193646)
--(axis cs:2048,98.2156829193646)
--(axis cs:1024,98.1296391871674)
--(axis cs:512,98.0560374328911)
--(axis cs:256,97.9221067636468)
--(axis cs:128,97.720963496859)
--(axis cs:64,97.3017862809814)
--(axis cs:32,95.4793168361686)
--(axis cs:16,92.1224054152866)
--(axis cs:8,83.195865430739)
--cycle;

\path [draw=shell, fill=shell, opacity=0.2]
(axis cs:8,76.6531559510697)
--(axis cs:8,4.68335167470799)
--(axis cs:16,15.2734432491719)
--(axis cs:32,27.4865109801099)
--(axis cs:64,34.4625187325866)
--(axis cs:128,40.8901498363918)
--(axis cs:256,47.1500091690119)
--(axis cs:512,51.4773862841786)
--(axis cs:1024,55.1030978810098)
--(axis cs:2048,57.7206647578217)
--(axis cs:2048,62.8891549053159)
--(axis cs:2048,62.8891549053159)
--(axis cs:1024,61.7173024814792)
--(axis cs:512,60.3544256166756)
--(axis cs:256,59.7391140696699)
--(axis cs:128,58.6235978336383)
--(axis cs:64,58.7873964932555)
--(axis cs:32,64.0326441277415)
--(axis cs:16,64.9992422978343)
--(axis cs:8,76.6531559510697)
--cycle;

\path [draw=quicksort, fill=quicksort, opacity=0.2]
(axis cs:8,57.025027512062)
--(axis cs:8,-14.9447767642998)
--(axis cs:16,-7.32919268203824)
--(axis cs:32,-1.41043150871498)
--(axis cs:64,3.7185760072967)
--(axis cs:128,8.21592281784305)
--(axis cs:256,11.876762419055)
--(axis cs:512,14.7821829969381)
--(axis cs:1024,17.3337088521218)
--(axis cs:2048,19.3042785034232)
--(axis cs:2048,45.0547995742391)
--(axis cs:2048,45.0547995742391)
--(axis cs:1024,46.1511140683053)
--(axis cs:512,46.9356452303429)
--(axis cs:256,48.3970431450205)
--(axis cs:128,48.8492564178331)
--(axis cs:64,52.0304860041808)
--(axis cs:32,54.6836333225334)
--(axis cs:16,55.9581879253503)
--(axis cs:8,57.025027512062)
--cycle;

\path [draw=corsort, fill=corsort, opacity=0.2]
(axis cs:8,11.2260611543772)
--(axis cs:8,-8.40206728463052)
--(axis cs:16,-5.06892908891723)
--(axis cs:32,-2.26034158191571)
--(axis cs:64,0.00227523830562681)
--(axis cs:128,1.51351727557663)
--(axis cs:256,2.90999430585053)
--(axis cs:512,3.89232660649115)
--(axis cs:1024,4.60706689673571)
--(axis cs:2048,5.13689543114171)
--(axis cs:2048,6.19919380137766)
--(axis cs:2048,6.19919380137766)
--(axis cs:1024,6.14658003650017)
--(axis cs:512,6.16319713340898)
--(axis cs:256,6.23541824187339)
--(axis cs:128,6.40068798347921)
--(axis cs:64,6.75918572738032)
--(axis cs:32,7.08866922329234)
--(axis cs:16,8.49265246980888)
--(axis cs:8,11.2260611543772)
--cycle;

\path [draw=merge, fill=merge, opacity=0.2]
(axis cs:8,11.2260611543772)
--(axis cs:8,-14.9447767642998)
--(axis cs:16,-5.06892908891723)
--(axis cs:32,-2.26034158191571)
--(axis cs:64,0.00227523830562681)
--(axis cs:128,1.09461692918498)
--(axis cs:256,1.42543004869746)
--(axis cs:512,1.62145607957329)
--(axis cs:1024,1.61927102548916)
--(axis cs:2048,1.58228165381369)
--(axis cs:2048,2.08278761671332)
--(axis cs:2048,2.08278761671332)
--(axis cs:1024,2.4175370979596)
--(axis cs:512,2.86011273061939)
--(axis cs:256,3.50382000871177)
--(axis cs:128,4.30618625152097)
--(axis cs:64,5.74564915401912)
--(axis cs:32,7.08866922329234)
--(axis cs:16,10.7529160629299)
--(axis cs:8,11.2260611543772)
--cycle;

\path [draw=binary, fill=binary, opacity=0.2]
(axis cs:8,11.2260611543772)
--(axis cs:8,-8.40206728463052)
--(axis cs:16,-5.06892908891723)
--(axis cs:32,-2.26034158191571)
--(axis cs:64,-1.01126133505559)
--(axis cs:128,-0.301717558787185)
--(axis cs:256,0.119013502402776)
--(axis cs:512,0.279578040940032)
--(axis cs:1024,0.364852911606994)
--(axis cs:2048,0.407624802110451)
--(axis cs:2048,0.770236265027546)
--(axis cs:2048,0.770236265027546)
--(axis cs:1024,0.935042963371613)
--(axis cs:512,1.20857052922461)
--(axis cs:256,1.54419518926971)
--(axis cs:128,2.21168451956273)
--(axis cs:64,3.04288495838925)
--(axis cs:32,4.53893900369016)
--(axis cs:16,8.49265246980888)
--(axis cs:8,11.2260611543772)
--cycle;

\path [draw=fj, fill=fj, opacity=0.2]
(axis cs:8,4.68335167470799)
--(axis cs:8,-8.40206728463052)
--(axis cs:16,-2.8086654957962)
--(axis cs:32,-2.26034158191571)
--(axis cs:64,-0.673415810601852)
--(axis cs:128,-0.441351007584401)
--(axis cs:256,-0.118516778741717)
--(axis cs:512,0.047329918868888)
--(axis cs:1024,0.136776890901147)
--(axis cs:2048,0.172693431769821)
--(axis cs:2048,0.41273200581351)
--(axis cs:2048,0.41273200581351)
--(axis cs:1024,0.513102325065806)
--(axis cs:512,0.666658244391938)
--(axis cs:256,0.890986916122372)
--(axis cs:128,1.23425037798219)
--(axis cs:64,1.69150286057429)
--(axis cs:32,2.83911885728869)
--(axis cs:16,3.97212528356685)
--(axis cs:8,4.68335167470799)
--cycle;

\addplot [dashed, semithick, heap, mark=*, mark size=1, mark options={solid}]
table {%
8 70.1104464714005
16 83.0813510428025
32 89.5299463237635
64 93.2476399875365
128 95.2075614185091
256 96.3187773659214
512 97.0238235570193
1024 97.491026329191
2048 97.801999419417
};
\addlegendentry{Heapsort}
\addplot [dotted, semithick, shell, mark=*, mark size=1, mark options={solid}]
table {%
8 37.3968990730542
16 37.8760791803821
32 42.7848922977231
64 45.2735755151061
128 49.1285233154276
256 53.028883627338
512 55.7868792159431
1024 58.2847583698563
2048 60.1823369426955
};
\addlegendentry{Shellsort}
\addplot [semithick, quicksort, mark=*, mark size=1, mark options={solid}]
table {%
8 4.68335167470799
16 13.0131796560509
32 16.4376800285004
64 19.935161181076
128 22.3189011463619
256 24.6440150305714
512 26.3429784067016
1024 27.7681867994142
2048 28.8700710392499
};
\addlegendentry{Quicksort}
\addplot [semithick, corsort, mark=*, mark size=1, mark options={solid}]
table {%
8 4.68335167470799
16 1.71186169044584
32 2.83911885728869
64 3.38073048284298
128 4.02691935392654
256 4.57270627386197
512 5.00195652305326
1024 5.39392916817087
2048 5.66804461625969
};
\addlegendentry{Corsort-$\Delta$}
\addplot [semithick, merge, mark=*, mark size=1, mark options={solid}]
table {%
8 4.68335167470799
16 3.97212528356685
32 3.68902893048941
64 3.04288495838925
128 2.77021831475159
256 2.49431631384767
512 2.26658975199314
1024 2.02980786275966
2048 1.83764183896655
};
\addlegendentry{Mergesort}
\addplot [semithick, binary, mark=*, mark size=1, mark options={solid}]
table {%
8 4.68335167470799
16 1.71186169044584
32 1.13929871088723
64 1.01581181166683
128 0.954983480387761
256 0.831604345836245
512 0.744074285082319
1024 0.661351738524596
2048 0.591484135420539
};
\addlegendentry{Binary insertion}
\addplot [semithick, fj, mark=*, mark size=1, mark options={solid}]
table {%
8 4.68335167470799
16 1.71186169044584
32 1.13929871088723
64 0.677966287213083
128 0.536083133996113
256 0.415926353833385
512 0.382799428527214
1024 0.330641508501128
2048 0.295266320643206
};
\addlegendentry{Ford-Johnson}
\end{axis}

\end{tikzpicture}
	\vspace{-.9cm}
	\caption{
	Number of comparisons required for algorithm termination, as a relative overhead compared to the information-theoretic lower bound. Each point is obtained by sorting $10,000$ random lists.}
	\label{fig:total}
	\vspace{-.3cm}
\end{figure}
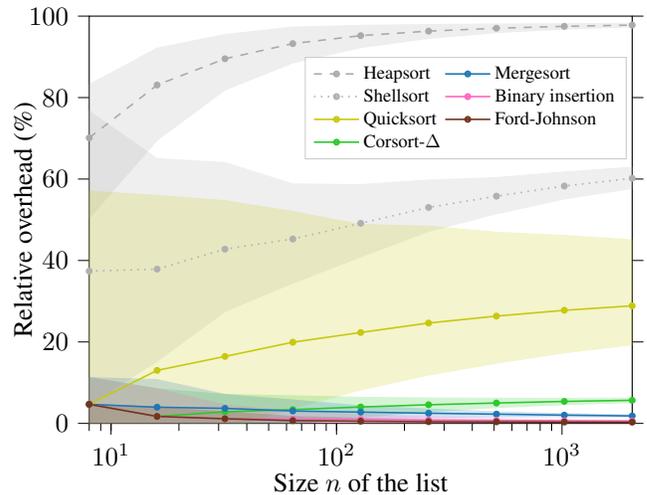

Figure \ref{fig:total} shows the termination time (measured in number of comparisons) for the different algorithms. The curves for top-down merge, bottom-up merge, and multizip sort coincide as they represent different scheduling variations of mergesort. Similarly, the curves for ASort and quicksort are identical as they involve the same comparisons. We consider values of the list size $n$ ranging from $8$ to $2048$. For each point, we generate 10,000 random lists and we record the number of comparisons required for a complete sort. The y-axis shows the relative deviation from the theoretical lower bound of $\log_2(n!) = n \log_2(n) - n/\ln(2) + \log_2(2\pi n)/2 + o(1)$~\cite{cormen}: a curve closer to $0$ indicates closer proximity to the optimal performance. To give a comprehensive view of the results distribution, each algorithm's median is depicted by a dark curve, while the 2.5\% to 97.5\% quantiles form a 95\% confidence interval represented by a light area.

We observe that Heapsort requires almost twice as many comparisons as necessary, and Shellsort (here implemented with Ciura's gap sequence~\cite{ciura2001best}) approximately 60\% more.
Quicksort shows better performance with less than 30\% overhead for the studied values of~$n$, but relying on a pivot selection introduces high variance.
The remaining four sorting algorithms demonstrate even lower overhead: 6\% for Corsort, 2\% for mergesort, 0.6\% for binary insertion, and $0.3\%$ for Ford-Johnson. These algorithms also exhibit negligible variance. Based on these findings, we conclude that Corsort emerges as a promising candidate, surpassing all other algorithms except those with comparison complexities asymptotically equivalent to $n\log_2(n)$~\cite{Ford1959ATP}. Moreover, our simulations confirm the asymptotical optimality of mergesort, further reinforcing the position of its multizip variant as a strong candidate for anytime sorting.

\subsection{Performance profiles}
\label{sec:profiles}

\begin{figure*}[!thb]
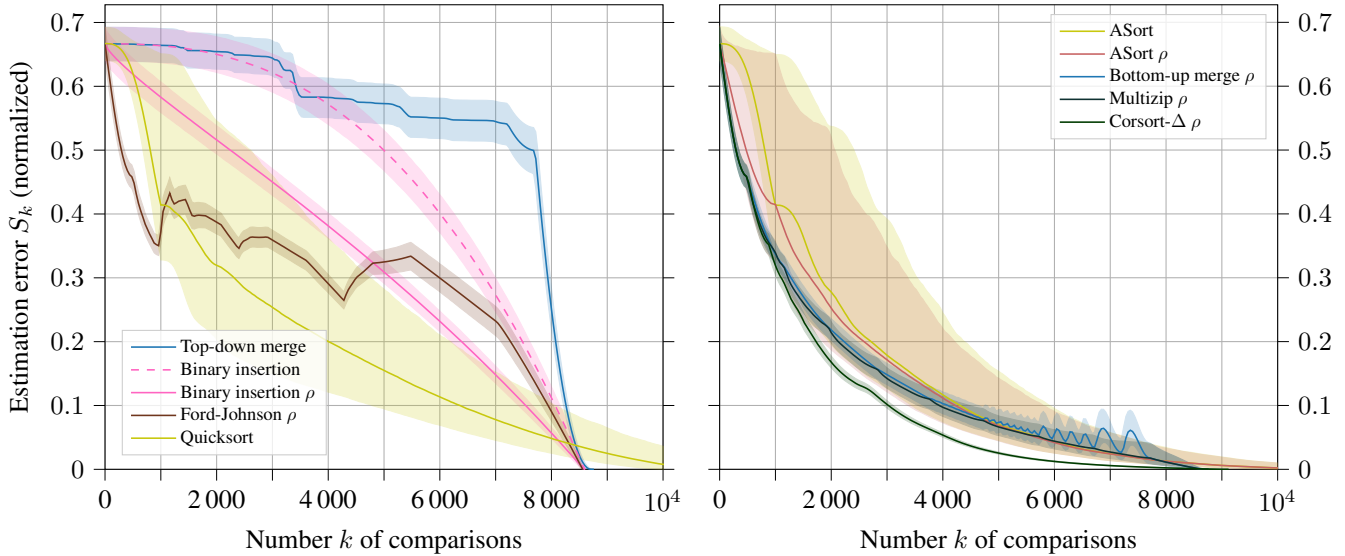

\include{bad_profiles}
	\vspace{-.5cm}
	\caption{Performance profiles for $n=1000$. Each curve is obtained by sorting $10,000$ random lists. The error $S_k$ is normalized by the maximal distance $\lfloor n^2/2\rfloor$. Results are dispatched on two subfigures for better visibility. For the computation times, cf. Appendix~\ref{sec:computation_time}\cfcompanion.
 }
	\label{fig:traj}
	\vspace{-.4cm}
\end{figure*}

Figure \ref{fig:traj} displays the performance profiles of the sorting algorithms. Heapsort and Shellsort were excluded due to their poor termination times (cf. \Cref{sec:termination_time}).
The Ford-Johnson algorithm employs the estimator $\rho$. 
Quicksort is examined in three variants: its natural implementation; ASort; ASort version equipped with~$\rho$. 
Three variants of mergesort are also considered: the natural top-down implementation; bottom-up merge; multizip sort. The last two are both equipped with~$\rho$. Binary insertion sort is explored with both its natural estimator and the estimator $\rho$. Finally, as discussed in \Cref{sec_Corsort}, Corsort utilizes $\Delta$ for selecting the next comparison (with additional criteria in case of ties) and $\rho$ for yielding estimates.

For each algorithm, we sort 10,000 lists of size $n=1000$. After each specified number of comparisons~$k$, we interrupt the algorithm and compute the estimation error $S_k$. As for \Cref{fig:total}, we represent the median and a 95\% confidence interval. The error is normalized by 
 the maximal distance $\lfloor n^2/2\rfloor$. The bounds from \cite{giesen2009approximate} are not displayed because they are irrelevant for the considered parameters: the lower bound starts from 0.005 and rapidly vanishes, while the ASort guaranteed upper bound is always above 1.

First and foremost, Corsort stands out with its remarkable performance profile. It consistently exhibits a decreasing trend with minimal variance, and outperforms other algorithms except for termination. 
Corsort emerges as a highly commendable anytime sorting algorithm, providing accurate estimates within a reasonable timeframe.

Multizip sort demonstrates notable performance, ranking second almost constantly in terms of the performance profile. It offers a compelling option in scenarios where global speed is of utmost importance, with a termination cost of $O(n\ln(n))$ total operations (excluding the interruption cost).

ASort also exhibits a good profile, especially if one focuses on the median values, for which it is slightly better than multizip for $k\in[5700, 7700]$. Its main drawback is a significant variance, shared with Quicksort, primarily explained by their pivot selection process. In contrast, the other sorting algorithms showcase greater consistency in their performance: binary insertion sorts, mergesorts, and Ford-Johnson exhibit relatively low variances, while Corsort demonstrates virtually negligible variance.

Corsort, multizip, and ASort are anytime by design and outperform algorithms such as binary insertion sort and Ford-Johnson. Yet, the differences between them are also significant, as illustrated by Table~\ref{tab:targets}, which links the number of comparisons and the estimation  error on a few values.
\begin{table}[ht]
    \centering
    \begin{tabular}{cccc}
    \hline
     \# of comparisons & Corsort & Multizip & ASort \\
     \hline
    4000 & 5.4 (5.7) & {}~~9.7 (11.4) & 11.7 (24.4)\\
    6000 & 1.2 (1.3) & 4.4 (6.2) & 4.3 (9.0)\\
    8000 & 0.2 (0.2) & 1.1 (2.0) & 1.3 (3.4)\\
    \hline
    \end{tabular}
    \caption{
    Normalized error (in percent) depending on the number of comparisons for $n=1000$. Each entry shows the median value, then the 97.5\% quantile in parenthesis. Numbers are rounded to the first decimal place.
    \label{tab:targets}
    }
\end{table}

We also observe that the use of the estimator~$\rho$ contributes to an improved performance profile. Figure~\ref{fig:traj} only shows the case of binary insertion and ASort to limit the number of curves, but we actually observed the same phenomenon for all algorithms from Table~\ref{tab:complexity} that have a native estimator. However, the impact of $\rho$ is limited for ASort, where the gain is relatively small and limited to the left part of the profile.

For bottom-up merge, $\rho$ induces a non-monotonic behavior towards the later stages of the sorting process. Indeed, $\rho$ is optimal when the partial order is a set of independent chains, but $\rho$ is inaccurate when there are also (non degenerated) Y-shapes, like in Figure~\ref{fig:differences_estimators}. Near completion of the bottom-merge algorithm, the partial orders fluctuate between these two configurations, which leads to the observed phenomenon.
For example, the final two ``bumps'' in Figure~\ref{fig:traj} correspond to the merge of quarters of the list. In contrast, by performing a round-robin on merge operations, our improved algorithm multizip stays monotonic and remains consistently below bottom-up merge equipped with $\rho$.

\section{Conclusion}\label{sec_conclu}

We have formalized the problem of anytime sorting and introduced novel approaches to address it. Our proposed estimators allow to transform any comparison-based sorting algorithm into an anytime algorithm. We have introduced the innovative algorithms of multizip sort and Corsort. Through extensive simulations, their outstanding performance has been demonstrated, showcasing their efficiency in terms of termination time and the quality of intermediate estimates. Our work advances the understanding and effectiveness of anytime sorting algorithms. In particular, despite its lack of proven upper bound and global complexity, we believe that Corsort \emph{de facto} replaces ASort as a benchmark for empirical evaluation of future anytime sorting algorithms.

The relative novelty of our approach opens up several avenues for future research. Exploring different scoring functions could potentially enhance Corsort even further, and also improve the performance profiles of classical sorting algorithms like Ford-Johnson. Formally proving that Corsort has an $O(n\ln(n))$ comparison complexity poses another challenge. Additionally, narrowing the gap between existing theoretical bounds and empirical observations is a promising direction for future research.

\paragraph{Acknowledgment} The work presented in this paper has been carried out at LINCS (\url{https://www.lincs.fr/}).
\newpage

\bibliographystyle{named}
\bibliography{corsort}

\newpage

\begin{myAppendix}

\section{Details of presented algorithms}

\subsection{Multizip sort}\label{sec:pseudocode_multizip}\label{sec:detailed_execution_multizip}

For the sake of clarity and conciseness, we present the pseudocode of multizip sort (\Cref{alg:multizip_sort}) in a version that does not sort the list in place. It uses two subalgorithms:
\begin{itemize}[align=left, leftmargin=\parindent, labelindent=\parindent,
 listparindent=\parindent, labelwidth=0pt, itemindent=!]
	\item Split (\Cref{alg:split}) takes several lists and cuts each of them into two lists of equal sizes (up to one element). In multizip sort, it is called enough times so that all lists end up being of size 0 or 1; in particular, all of them are trivially sorted.
	\item Multizip merge (Algorithm~\ref{alg:multizip_merge}) takes an even number of sorted lists $Y_1, \ldots, Y_{2\ell}$ and merges them two by two, as in the classical bottom-up merge sort. The particularity here is that instead of completely merging $Y_1$ and $Y_2$ (``closing one zipper''), then $Y_3$ and $Y_4$ (``closing another zipper''), etc, we perform only the first comparison for each merge operation, then only the second comparison for each merge operation, etc. (metaphorically closing all zippers in parallel, little by little, in a round-robin fashion). 
\end{itemize}

\begin{algorithm}[H] 
	\caption{Multizip sort}
	\label{alg:multizip_sort}
	\textbf{Input}: List $X$.\\
	\textbf{Output}: List $X$ sorted in ascending order.\\[-\baselineskip]
	\begin{algorithmic}[1] 
		\State $Y_1 := X$.
		\State $\ell := 1$.
		\While {$\exists i \in \{1, \ldots, \ell\}, |Y_i| > 1$}
		\State $Y_1, \ldots, Y_{2\ell}$ := Split($Y_1, \ldots, Y_\ell$).
		\State $\ell := 2\ell$.
		\EndWhile
		\While {$\ell > 1$}
		\State $Y_1, \ldots, Y_{\ell/2}$ := Multizip merge($Y_1, \ldots, Y_\ell$).
		\State $\ell := \ell / 2$.
		\EndWhile
		\Return $Y_1$.
	\end{algorithmic}
\end{algorithm}

\begin{algorithm}[H] 
	\caption{Split}
	\label{alg:split}
	\textbf{Input}: Lists $Y_1, \ldots, Y_\ell$.\\
	\textbf{Output}: Lists $Z_1, \ldots, Z_{2\ell}$, where for each $i \in \{1, \ldots, \ell\}$, $Y_i = \text{concatenate}(Z_{2i-1}, Z_{2i})$, $|Z_{2i-1}| = \lceil |Y_i| / 2 \rceil$, and $|Z_{2i}| = \lfloor |Y_i| / 2 \rfloor$.\\[-\baselineskip]
	\begin{algorithmic}[1] 
		\For {$i = 1$ to $\ell$}
		\State $y_i := |Y_i|$.
		\State $m_i := \lceil y_i / 2 \rceil$.
		\State $Z_{2i-1} := \big[Y_i[1], \ldots, Y_i[m_i]\big]$.
		\State $Z_{2i} := \big[Y_i[m_i + 1], \ldots, Y_i[y_i]\big]$.
		\EndFor
		\Return $Z_1, \ldots, Z_{2\ell}$.
	\end{algorithmic}
\end{algorithm}

\begin{algorithm}[H] 
	\caption{Multizip merge}
	\label{alg:multizip_merge}
	\textbf{Input}: An even number of lists $Y_1, \ldots, Y_{2\ell}$, where each list $Y_i$ is sorted in ascending order.\\
	\textbf{Output}: Lists $Z_1, \ldots, Z_\ell$, where each list $Z_i$ consists of the elements of $Y_{2i-1}$ and $Y_{2i}$, sorted in ascending order.\\[-\baselineskip]
	\begin{algorithmic}[1] 
		\State $Z_i := [], \; \forall i \in \{1, \ldots, \ell\}$.
		\While{$\exists i, Y_i \neq []$}
		\For{$i = 1$ to $\ell$}
		\If {$Y_{2i-1} = []$ or $Y_{2i} = []$}
		\State $Z_i$ = concatenate($Z_i, Y_{2i-1}, Y_{2i}$).
		\State Empty $Y_{2i-1}$ and $Y_{2i}$.
		\Else
		\If {$Y_{2i-1}[1] \leq Y_{2i}[1]$}
		\State $Z_i$.append($Y_{2i-1}$.pop(1)).
		\Else
		\State $Z_i$.append($Y_{2i}$.pop(1)).
		\EndIf
		\EndIf
		\EndFor
		\EndWhile
		\Return $Z_1, \ldots, Z_\ell$.
	\end{algorithmic}
\end{algorithm}

In \Cref{fig:multizip_exec}, we presented the execution of multizip sort on the list $(51872643)$, which we recall in \Cref{fig:onlymultizipexecutions} for ease of reading. Here we detail this example. 

\begin{figure}[H]
\begin{center}
\begin{tikzpicture}[scale=.53, transform shape]
	\def\interspace{.07cm}
	\node (x0) {\execMerge{5, 1, 8, 7, 2, 6, 4, 3}{5/1, 8/7, 2/6, 4/3}{0/1/2, 2/3/4, 4/5/6, 6/7/8}};
	\node[below = \interspace of x0] (x1) {\execMerge{1, 5, 7, 8, 2, 6, 3, 4}{1/7, 2/3}{0/2/4, 4/6/8}};
	 \node[below = \interspace of x1] (x2) {\execMerge{1, 5, 7, 8, 2, 6, 3, 4}{5/7, 6/3}{1/2/4, 5/6/8}};
	\node[below = \interspace of x2] (x3) {\execMerge{1, 5, 7, 8, 2, 3, 6, 4}{6/4}{1.88/2.08/4, 6/7/8}};
	\node[below = \interspace of x3] (x4) {\execMerge{1, 5, 7, 8, 2, 3, 4, 6}{1/2, 5/2, 5/3, 5/4, 5/6, 7/6}{0/4/8}};
	\node[below = \interspace of x4] (x5) {\execMerge{1, 2, 3, 4, 5, 6, 7, 8}{}{}};
\end{tikzpicture}    
\end{center}
	\caption{Sorting the list $X=(51872643)$ with multizip sort. Each edge represents a comparison. The bracket notation $[\;\mid\;]$ delimits two sublists already sorted that are being merged.\label{fig:onlymultizipexecutions}}
	\vspace{-.1cm}
\end{figure}
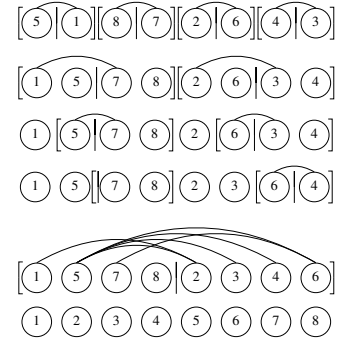

After enough splitting operations, the initial list is decomposed into sublists of size 0 or 1. Since, for simplicity, we chose an example whose size is a power of two, the decomposition contains only sublists of size 1 and no empty list. Then the merging process begins (the line numbers below refer to Figure~\ref{fig:multizip_exec} or~\ref{fig:onlymultizipexecutions}).
\begin{itemize}[align=left, leftmargin=\parindent, labelindent=\parindent,
 listparindent=\parindent, labelwidth=0pt, itemindent=!]
    \item First multizip merge:
    \begin{itemize}
        \item Line 1: $Y_1 = [5]$, $Y_2 = [1]$, $Y_3 = [8]$, $Y_4 = [7]$, $Y_5 = [2]$, $Y_6 = [6]$, $Y_7 = [4]$, $Y_8 = [3]$. Compare $(5, 1)$, $(8, 7)$, $(2, 6)$, and $(4, 3)$, like in bottom-up merge sort. 
    \end{itemize}
    \item Second multizip merge:
    \begin{itemize}
        \item Line 2: $Y_1 = [1, 5]$, $Y_2 = [7, 8]$, $Y_3 = [2, 6]$, $Y_4 = [3, 4]$. Compare $(1, 7)$ and $(2, 3)$.
        \item Line 3: $Z_1 = [1]$, $Y_1 = [5]$, $Y_2 = [7, 8]$, $Z_2 = [2]$, $Y_3 = [6]$, $Y_4 = [3, 4]$. Compare $(5, 7)$ and $(6, 3)$.
        \item Line 4: $Z_1 = [1, 5]$, $Y_1 = []$, $Y_2 = [7, 8]$, $Z_2 = [2, 3]$, $Y_3 = [6]$, $Y_4 = [4]$. For $Z_1$, no more comparaison is needed because $Y_1$ is empty. Compare $(6, 4)$ to finish merging $Y_3$ and $Y_4$ into $Z_2$.
    \end{itemize}
    \item Third multizip merge:
    \begin{itemize}
        \item Line 5: $Y_1 = [1, 5, 7, 8]$, $Y_2 = [2, 3, 4, 6]$. Compare $(1, 2)$, $(5, 2)$, $(5, 3)$, $(5, 4)$, $(5, 6)$, and $(7, 6)$, like in top-down or bottom-up merge sort.
    \end{itemize}
\end{itemize}

Note that in multizip sort, the first multizip merge is always identical to the first round of merging operations in bottom-up merge sort. Indeed, each sublist is of size 0 or 1, hence each merging operation involves at most one comparison, which implies that there is no difference between treating each merging operation as a block and performing the necessary comparisons in a round-robin fashion. On the other hand, the last merging operation (between the two sorted half-lists) is always the same for the three variants: top-down merge sort, bottom-up merge sort, and multizip sort. Hence to show the difference between bottom-up merge sort and multizip sort, it is necessary to have an example with three levels of merging, such as the one we present.

\subsection{Partial orders and estimators}
\label{sec:pseudocode_estimators}

\Cref{alg:partialorderupdate} (Partial order update) performs the update of a matrix $M$ representing a partial order, when we acquire the information of a new comparison~$i \prec j$. In the paper, it is used in two contexts.
\begin{itemize}[align=left, leftmargin=\parindent, labelindent=\parindent,
 listparindent=\parindent, labelwidth=0pt, itemindent=!]
    \item Corsort calls \Cref{alg:partialorderupdate} after each comparison  and uses the updated partial order to choose the next pair to compare.
    \item The anytime adaptation of any sorting algorithm, defined in \Cref{subsec:estimators}, builds the partial order only in case of interruption, by iterating \Cref{alg:partialorderupdate} for every comparison in the history.
\end{itemize}

\begin{algorithm}[H]
    \caption{Partial order update}
    \label{alg:partialorderupdate}
    \textbf{Input}: A matrix $M$ such that $M_{k,\ell}=+1$ if $k \preceq \ell$, $M_{k,\ell}=-1$ if $k \succ \ell$, and $M_{k,\ell}=0$ if we have not compared $k$ and $\ell$ yet. A pair $(i, j)$. \\
    \textbf{Output}: The transitive closure of $M$ with the additional comparison $i\preceq j$. \\[-\baselineskip]
    \begin{algorithmic}[1] 
        \State descendants$_i := \{ k, M_{k,i}=+1\}$.
        \State ascendants$_j := \{ \ell, M_{j,\ell}= +1\}$.
        \For{$k \in \text{descendants}_i$}
            \For{$\ell \in \text{ascendants}_j$}
                \State $M_{k,\ell} := +1$.
                \State $M_{\ell,k} := -1$.
            \EndFor
        \EndFor
		\Return $M$.
	\end{algorithmic}
\end{algorithm}
Let $M$ be defined with the convention of \Cref{alg:partialorderupdate} and $i$ be the index of an element. To compute the number of ascendants $a(i)$ or descendants $d(i)$, it suffices to read the row or column $i$ of matrix $M$:
\begin{align*}
    a(i) &= \sum_{j=1}^{n} \mathbb{1}[ M_{i,j}=+1], \\
    d(i) &= \sum_{j=1}^{n} \mathbb{1}[ M_{j,i}=+1].
\end{align*}

The computation of $I(i)$, $\Delta(i)$, and $\rho(i)$ is straightforward:
\begin{align*}
    I(i) &= d(i) +a(i), \\
    \Delta(i) &= d(i) - a(i), \\
    \rho(i) &= \frac{d(i)}{d(i)+a(i)}.
\end{align*}

\subsection{Corsort}\label{sec:pseudocode_corsort}
\Cref{alg:corsort} describes Corsort-$\Delta$ $\rho$, which uses $\Delta$ to choose the next comparison to perform and $\rho$ to estimate the sorted list. Some variants can be obtained by replacing $\Delta$ and $\rho$ with other estimators.
A detailed description of Corsort is available in \Cref{sec_Corsort}.
\begin{algorithm}[H] 
	\caption{Corsort-$\Delta$ $\rho$}
	\label{alg:corsort}
	\textbf{Input}: List $X$ of size $n$. \\
	\textbf{Output}: List $X$ sorted in ascending order if the algorithm ends with no interruption. An estimate of the sorted list otherwise. \\[-\baselineskip]
	\begin{algorithmic}[1] 
		\State $M := $ the identity matrix of
        size $n$, representing the empty partial order on $X$.
		\While{there is no interruption and $\exists (i,j), \; M_{i,j}=0$ }
        \State Update $\Delta$ and $I$ according to $M$.
        \State $\mathcal{S}:= \{(i, j), \; M_{i,j}=0 \}$.
		\State  $(i, j) := \argmin_{\mathcal{S}} \big(|\Delta(i) - \Delta(j)|, \max(I(i), I(j))\big)$.
        \If{$X[i] > X[j]$}
            \State $(i, j) := (j, i)$.
        \EndIf
		\State $M :=$ Partial order update$\big(M, (i, j)\big)$.
		\EndWhile
        \State  Update $\rho$ according to $M$. \\
		\Return $X$ sorted by ascending value of $\rho$.
	\end{algorithmic}
\end{algorithm}

\section{Implementation}\label{sec:implementation}

\subsection{The Corsort package}

The Python code of this article has been published on \href{https://github.com/emczg/corsort}{GitHub}\footnote{\url{https://github.com/emczg/corsort}} and \href{https://pypi.org/project/corsort/}{PyPI}\footnote{\url{https://pypi.org/project/corsort/}} under GNU General Public license v3~\cite{corsort}. The \href{https://emczg.github.io/corsort/}{documentation}\footnote{\url{https://emczg.github.io/corsort/}} provides detailed instructions for package installation and for the replication of our results.

In particular, the \emph{Notebooks} section of the documentation gives step-by-step intructions to reproduce all the experiments we describe.
\begin{itemize}[align=left, leftmargin=\parindent, labelindent=\parindent,
 listparindent=\parindent, labelwidth=0pt, itemindent=!]
    \item The \emph{\href{https://emczg.github.io/corsort/notebooks/termination.html}{Termination times}}\footnote{\url{https://emczg.github.io/corsort/notebooks/termination.html}} notebook explains the computation of Figure~\ref{fig:total}.
    \item The \emph{\href{https://emczg.github.io/corsort/notebooks/corsorts.html}{Comparison of Corsort variants}}\footnote{\url{https://emczg.github.io/corsort/notebooks/corsorts.html}} notebook explains our selection of the Corsort-$\Delta$ $\rho$ variant.
    \item The \emph{\href{https://emczg.github.io/corsort/notebooks/rho_delta.html}{Impact of estimators}}\footnote{\url{https://emczg.github.io/corsort/notebooks/rho_delta.html}} notebook compares natural estimators (if they exist) and the external estimators $\Delta$ and~$\rho$.
    \item The \emph{\href{https://emczg.github.io/corsort/notebooks/profiles.html}{Performance profiles}}\footnote{\url{https://emczg.github.io/corsort/notebooks/profiles.html}} notebook explains the computation of Figure~\ref{fig:traj}.
    \item The \emph{\href{https://emczg.github.io/corsort/notebooks/examples_with_a_chain_and_a_y_shape.html}{Examples with a chain and a Y-shape}}\footnote{\url{https://emczg.github.io/corsort/notebooks/examples_with_a_chain_and_a_y_shape.html}} notebook gives toy examples that illustrate the behavior of estimators. In particular, it explains how the example of Figure~\ref{fig:differences_estimators} was built.
\end{itemize}

\subsection{Running time}\label{sec:computation_time}

For our benchmark of Figure~\ref{fig:traj}, we sort $10,000$ random lists of size $1,000$. We choose this size of list for two reasons. First, our framework addresses applications where the comparisons are expensive, for example because they are manually performed by human experts, so the typical use-case does not concern lists whose size is orders of magnitude higher. Secondly, since this benchmark consists in interrupting the algorithms at each step to check the current Spearman distance to the ground truth, it slows down the process; without this artificial additional delay, it is realistic to use the algorithms with larger lists.

In the table below, we give the running time for each algorithm. The hardware used was a AMD Ryzen Threadripper 1950X (32 cores). Note that actual running time (i.e. in seconds or minutes) is not the primary focus of the paper.

\begin{center}
	\begin{tabular}{|c|r|}
		\hline
		Algorithm & \multicolumn{1}{c|}{Time} \\ \hline
		Top-down merge & 10 min 03 s \\
		Binary insertion & 8 min 21 s \\
		Binary insertion $\rho$ & 21 min 02 s \\
		Ford-Johnson $\rho$ & 24 min 52 s \\
		Quicksort  & 10 min 13 s \\
		ASort  & 10 min 33 s \\
		ASort $\rho$ & 39 min 56 s \\
		Bottom-up merge $\rho$  & 16 min 39 s \\
		Multizip $\rho$  & 16 min 42 s \\
		Corsort-$\Delta$ $\rho$  & 78 min 35 s \\
		\hline
	\end{tabular}
\end{center}

\end{myAppendix}

\end{document}